\documentclass[twocolumn,showpacs,aps,prd,floatfix]{revtex4}

\usepackage{graphicx}
\usepackage{dcolumn}
\usepackage{epsfig}
\usepackage{psfrag}
\usepackage{amsmath}
\usepackage{longtable}
\usepackage{bm}
 
\long\def\inst#1{\par\nobreak\kern 4pt\nobreak
    {\it #1}\par\vskip 10pt plus 3pt minus 3pt}
\usepackage{relsize}
\usepackage{xspace}
\def\babar{\mbox{\slshape B\kern-0.1em{\smaller A}\kern-0.1em
    B\kern-0.1em{\smaller A\kern-0.2em R}}}
\def\pep2{PEP-II}
\def\Kp    {\ensuremath{K^+}\xspace}
\def\Km    {\ensuremath{K^-}\xspace}

\def\pip   {\ensuremath{\pi^+}\xspace}
\def\pim   {\ensuremath{\pi^-}\xspace}
\def\Dz{D^0}

\def\D       {\ensuremath{D}\xspace}
\def\Dp      {\ensuremath{D^+}\xspace}
\def\Dm      {\ensuremath{D^-}\xspace}
\def\Dz      {\ensuremath{D^0}\xspace}
\def\Dzb     {\ensuremath{\Dbar^0}\xspace}
\def\DzDzb   {\ensuremath{\Dz {\kern -0.16em \Dzb}}\xspace}
\def\DpDm    {\ensuremath{\Dp {\kern -0.16em \Dm}}\xspace}
\def\Dstar   {\ensuremath{D^*}\xspace}
\def\Dstarb  {\ensuremath{\Dbar^*}\xspace}

\def\BR         {{\ensuremath{\cal B}\xspace}}
\def\invfb   {\ensuremath{\mbox{\,fb}^{-1}}\xspace}
\def\Dbar    {\kern 0.2em\overline{\kern -0.2em D}{}\xspace}
\def\Db      {\ensuremath{\Dbar}\xspace}
\def\DDb     {\ensuremath{\D {\kern -0.16em \Db}}\xspace}
\def\DDbX     {\ensuremath{\D {\kern -0.16em \Db}X}\xspace}
\def\DDstarb     {\ensuremath{\D {\kern -0.16em \Dstarb}}\xspace}
\def\DstarDstarb     {\ensuremath{\Dstar {\kern -0.16em \Dstarb}}\xspace}
\def\Y#1S{\ensuremath{\Upsilon{(#1S)}}\xspace}

\def\Ds      {\ensuremath{D^+_s}\xspace}

\def\pmpm      {\hspace{-0.18em}\pm\hspace{-0.18em}}

\newcommand{\gevc}{\ensuremath{{\mathrm{\,Ge\kern -0.1em V\!/}c}}\xspace}
\newcommand{\mevc}{\ensuremath{{\mathrm{\,Me\kern -0.1em V\!/}c}}\xspace}
\newcommand{\gevcc}{\ensuremath{{\mathrm{\,Ge\kern -0.1em V\!/}c^2}}\xspace}
\newcommand{\mevcc}{\ensuremath{{\mathrm{\,Me\kern -0.1em V\!/}c^2}}\xspace}
\newcommand{\mev}{\ensuremath{\mathrm{\,Me\kern -0.1em V\!}}\xspace}
\newcommand{\gevccc}{\ensuremath{{\mathrm{\,Ge\kern -0.1em V^2\!/}c^4}}\xspace}

\newcommand{\BaBarYear}    {08}
\newcommand{\BaBarNumber}  {049}
\newcommand{\SLACPubNumber} {13323}
\newcommand{\BaBarType}      {PUB}  
\begin{document}
\begin{flushleft}
\babar-\BaBarType-\BaBarYear/\BaBarNumber \\
SLAC-PUB-\SLACPubNumber \\
\end{flushleft}
 
\title{ 
\Large \bf\boldmath 
Dalitz Plot Analysis of
   \boldmath{$D_s^+ \to \pip \pim \pip$}
}

%
\author{B.~Aubert}
\author{M.~Bona}
\author{Y.~Karyotakis}
\author{J.~P.~Lees}
\author{V.~Poireau}
\author{E.~Prencipe}
\author{X.~Prudent}
\author{V.~Tisserand}
\affiliation{Laboratoire de Physique des Particules, IN2P3/CNRS et Universit\'e de Savoie, F-74941 Annecy-Le-Vieux, France }
\author{J.~Garra~Tico}
\author{E.~Grauges}
\affiliation{Universitat de Barcelona, Facultat de Fisica, Departament ECM, E-08028 Barcelona, Spain }
% ...
\collaboration{The \babar\ Collaboration}
\noaffiliation

\begin{abstract}
A Dalitz plot analysis of approximately $13,000$ \Ds
decays to $\pip \pim \pip$ has been performed. The analysis uses a   
384~${\rm fb}^{-1}$ data sample  
recorded by the \babar\  detector at the \pep2 asymmetric-energy $e^+e^-$
storage ring running at center of mass energies near 10.6 GeV.
Amplitudes and phases
of the intermediate resonances which contribute to this final state are measured.
A high precision measurement of the ratio of branching fractions is performed: 
$\BR(\Ds \to \pip \pim \pip)/\BR(\Ds \to \Kp \Km \pip)=0.199 \pm 0.004 
\pm 0.009$. Using a model-independent partial wave analysis, the amplitude and 
phase of the $\mathcal{S}$-wave have been measured.
\end{abstract}
\maketitle
\section{Introduction}\label{sec:intro}
Dalitz plot analysis is an excellent way to study
the dynamics of three-body charm decays.  
These decays are
expected to proceed predominantly through intermediate quasi-two-body modes~\cite{Bauer:1986bm} 
and experimentally
this is the observed pattern.
Dalitz plot analyses can provide new information on the resonances that
contribute to the observed three-body final states. 
In addition, since the intermediate quasi-two-body modes are dominated by
light quark meson resonances, new information on light meson spectroscopy 
can be obtained.

Some puzzles still remain in light meson spectroscopy. There are new 
claims for the existence of broad states close to threshold
such as $\kappa(800)$ and $f_0(600)$~\cite{e791}. 
The new evidence has reopened discussion of the composition of the ground state
$J^{PC}=0^{++}$ nonet, and of the possibility that states such as the 
$a_0(980)$
or $f_0(980)$ may be 4-quark states, due to their proximity to the
$K \bar K$ threshold~\cite{q4}. This hypothesis 
can be tested only through accurate measurements
of the branching fractions and the couplings to  
different final states. 
It is therefore important to have precise information 
on the structure of the $\pi \pi$ $\mathcal{S}$-wave. 
In addition, comparison between the production
of these states in decays of differently flavored charmed 
mesons $D^0 (c \bar u)$, $D^+(c \bar d)$ and $D_s^+(c \bar s)$ 
can yield new information on their 
possible quark 
composition. In this context, $D_s^+$ mesons can give information on 
the structure of the scalar amplitude coupled to $s \bar s$.
Another benefit of studying charm decays 
is that, in some cases, partial wave analyses
are able to isolate the scalar contribution 
with almost no background. 

This paper focuses on the study of the three-body 
\Ds meson decays to $\pip\pim\pip$~\cite{cc} and performs, for the first time, a 
Model-Independent Partial Wave Analysis (MIPWA)~\cite{Aitala:2005yh}.
Previous Dalitz plot analyses of this decay mode were based on much smaller 
data samples~\cite{e791_ds,focus} and did not have
sufficient statistics to perform the detailed analysis reported here.

This paper is organized as follows. Section II  briefly describes the 
\babar\ detector, while Section III gives details on event reconstruction.
Section IV is devoted to the evaluation of the selection efficiency. Section V
deals with the Dalitz plot analysis of $D_s^+ \to \pip\pim\pip$ and results are 
given in Section VI. The measurement of the $D_s^+$ branching fraction is described in Section VII.

\section{The \babar\ Detector and Dataset} \label{sec:detector}
Tha analysis is based on data collected with the \babar\ detector at the \pep2\ asymmetric-energy 
$e^+e^-$ collider at SLAC .  
The data sample used in this analysis corresponds to an integrated luminosity
of 347.5 \invfb recorded at the \Y4S resonance (on-peak) and 36.5 \invfb collected 40 \mev 
below the resonance (off-peak). 
The \babar\ detector is
described in detail elsewhere~\cite{Aubert:2001tu}.
The following is a brief summary of the 
components important to this analysis. 
Charged particles are detected
and their momenta measured by a combination of a cylindrical drift chamber (DCH)
and a silicon vertex tracker (SVT), both operating within a
1.5 T solenoidal magnetic field.
A ring-imaging Cherenkov detector (DIRC) is used for
charged-particle identification. Photon energies are measured with a
CsI electromagnetic calorimeter (EMC).
Information from the DIRC and energy-loss measurements in the DCH and 
SVT are used to identify charged kaon and pion candidates.
Monte Carlo (MC) events used in this analysis, $e^+ e^- \to c \bar c$, are generated 
using the JETSET program~\cite{jetset}, and the generated particles are propagated 
through a model of the \babar\ detector with the GEANT4 simulation package~\cite{geant4}.
Radiative corrections for signal and background processes are simulated using PHOTOS~\cite{photos}. 

\section{Event Selection and $D^+_s$ Reconstruction}
Events corresponding to the three-body decay:
\begin{eqnarray}
D^+_s \to \pip \pim \pip
\end{eqnarray}
are reconstructed from the sample of events having
at least three reconstructed charged tracks with a net charge of $\pm$1 and having
a minimum transverse momentum of 0.1 \gevc.
Tracks from $D^+_s$ decays are identified as pions or kaons by the Cherenkov
angle $\theta_c$ measured with the DIRC. The typical separation between pions and 
kaons varies from $8 \sigma$ at 2 \gevc to $2.5 \sigma$ at 4 \gevc, where $\sigma$ is
the average resolution on $\theta_c$. Lower momentum kaons are identified
with a combination of $\theta_c$ (for momenta down to 0.7 \gevc) and
measurements of ionization energy loss $dE/dx$ in the DCH and SVT.
The particle identification efficiency is $\approx$ 95\%,
while the misidentification rate for kaons is $\approx$ 5\%.
Photons are identified as EMC clusters that do not have a spatial match with a
charged track, and that have a minimum energy of 100 \mev. To reject background,
the lateral energy is required to be less than 0.8.
The three tracks are fitted to a common vertex, and the $\chi^2$ fit
probability (labeled $P_1$)
must be greater than 0.1~\%. A separate kinematic fit which makes use of the \Ds mass
constraint, to be used in the Dalitz plot analysis, is also performed. 
To help discriminate signal from background, 
an additional fit which uses
the constraint that the three tracks originate from the $e^+ e^-$
luminous region (beam spot) is performed. We label the $\chi^2$ probability of this fit as $P_2$,
and it is expected to be
large for background and small for \Ds signal events, since in general the latter will have a 
measurable flight distance.

The combinatorial background is reduced by requiring the $D^+_s$ to originate from the decay
\begin{eqnarray}
D^*_s(2112)^+ \to D^+_s \gamma
\end{eqnarray}
using the mass difference
$\Delta m=m(\pip \pim \pip \gamma)-m(\pip \pim \pip)$.
We cannot reliably extract the $D^*_s(2112)^+$ characteristics using the $3 \pi$ decay mode 
due to the large background below the signal peak.
Therefore we use the decay
\begin{eqnarray}
D^+_s \to \Kp \Km \pip,
\end{eqnarray}
which has a much larger signal to background ratio.
Fitting the mass difference $\Delta m=m(\Kp \Km \pip \gamma) - m(\Kp \Km \pip)$ for this decay mode 
with a polynomial describing the background and a single Gaussian for the signal,
we obtain a width $\sigma=5.51 \pm 0.04$ \mevcc. Since the experimental resolution in 
$\Delta m$ is similar for the two $\Ds$ decay modes, we 
require the value of $\Delta m$ for the $D^+_s \to \pip \pim \pip$ mode to be 
within $\pm 2\sigma$ of the Review of Particle Physics~\cite{pdg} value
of the ($D^*_s(2112)^+-D_s^+$) mass difference.
At this stage the three-pion invariant mass signal region, defined between
$(-2 \sigma,2 \sigma)$, where $\sigma$ is estimated by a Gaussian fit to the \Ds lineshapes,
has a purity (signal/(signal+background)) of 4.3\%. 

Each \Ds candidate is characterized by three
variables: the center of mass momentum $p^*$, the difference in probability
$P_1 - P_2$, and the signed decay distance $d_{xy}$ between the $D^+_s$ decay vertex
and the beam spot projected in the plane normal to the beam collision axis.
The distributions for these variables for background are inferred from the
$\Ds \to \pip \pim \pip$ invariant mass
sidebands defined between $(-9 \sigma,-5\sigma)$ 
and $(5 \sigma, 9 \sigma)$.  
Since these variables are (to a good approximation) independent of the decay mode, 
the distributions for the three-pion invariant mass signal, are inferred from the $\Ds \to \Kp \Km \pip$ decay.
These normalized distributions are then combined in a likelihood ratio test.
The cut on the likelihood ratio has been chosen in order to obtain the largest statistics
with background small enough to perform a Dalitz plot analysis.
 
Many possible background sources are examined. 
A small background contribution due to the decay $D^{*+} \to \pip D^0$ where $D^0 \to \pip \pim$ 
is addressed by removing events with $| m(\pip \pim)-m_{D^0} | < 20.7 $ \mevcc
and $m(\pip \pim \pip)-m(\pip \pim) < 0.1475 $ \gevcc.
Particle misidentification, in which a kaon ($K_{\rm mis}$) is wrongly identified as a pion,
is tested by assigning the kaon mass to each pion in turn. 
In this way we observe
a clean signal in the mass difference $m(\pip K_{\rm mis}^- \pip) - m(\pip K_{\rm mis}^-)$ 
due to the decay $D^{*+} \to \pip D^0$ where $D^0 \to K_{\rm mis}^- \pip$. Removing events with 
$| m(K_{\rm mis}^- \pip)-m_{D^0} | < 21.7 $ \mevcc
and $m(\pip K_{\rm mis}^- \pip)-m(K_{\rm mis}^- \pip) < 0.1475 $ \gevcc diminishes this background.
Finally, events having more than one candidate are removed from the sample 
(1.2 \% of the events).  

\begin{figure}[!htb]
\begin{center}
\includegraphics[height=7.8cm]{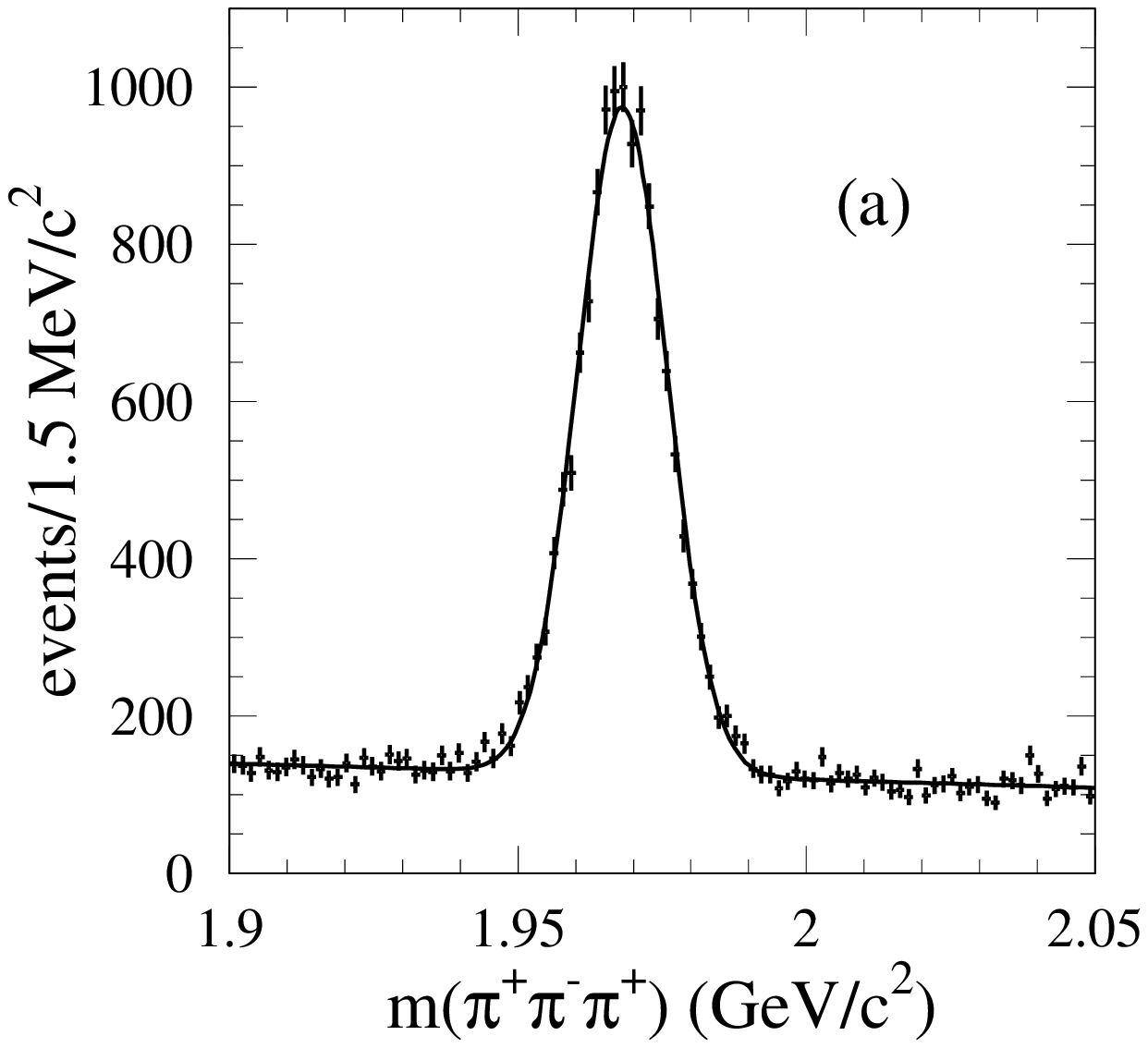}
\includegraphics[height=7.8cm]{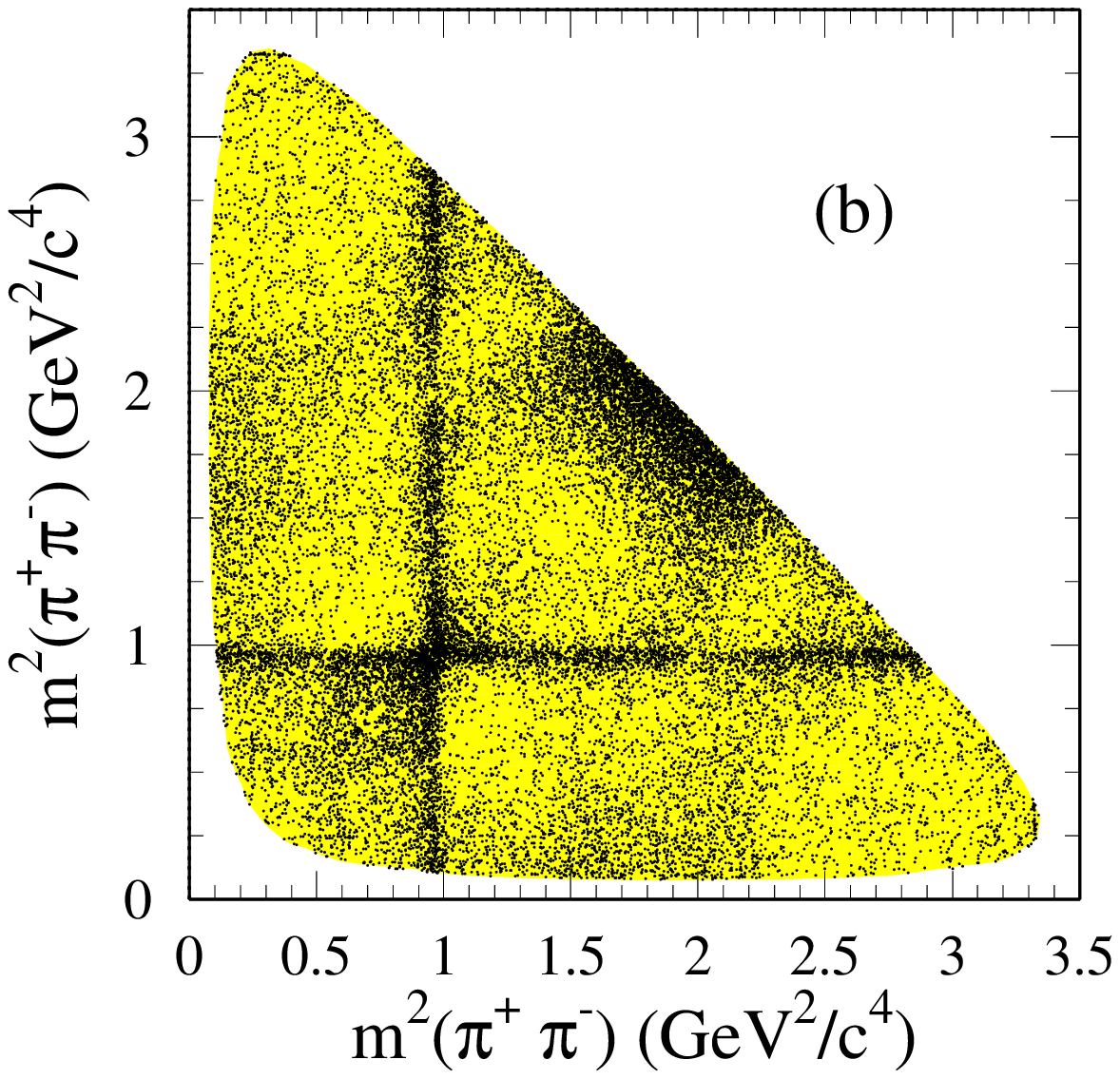}
\caption{(a) $\pip\pim\pip$ invariant mass distribution for the \Ds analysis sample.
The line is the result of the fit described in the text. (b) Symmetrized $D^+_s \to \pip \pim \pip$ Dalitz plot (two entries per event).}
\label{fig:fig_1}
\end{center}
\end{figure}

The resulting $\pip\pim\pip$ mass distribution is shown
in Fig.~\ref{fig:fig_1}(a). 
This distribution has been fitted with a single Gaussian
for the signal and a linear background function. The fit gives a $D_s^+$ mass
of  (1968.1 $\pm$ 0.1) \mevcc and width $\sigma=7.77 \pm 0.09$ \mevcc (statistical error only). 
The signal region contains 13179 events with a purity of 80\%.
The resulting Dalitz plot, symmetrized along the two axes, is shown 
in Fig.~\ref{fig:fig_1}(b). 
For this distribution, and in the following Dalitz plot 
analysis, we use the track momenta obtained from the \Ds mass-constrained fit.  
We observe a clear $f_0(980)$ signal, evidenced by the two narrow crossing bands.
We also observe a broad accumulation of events in the 
1.9 \gevccc region.

\section{Efficiency}
The efficiency for this \Ds decay mode is
determined from a sample of Monte Carlo events in which the \Ds
decay is generated
according to phase space ({\it i.e.} such that the Dalitz plot is uniformly
populated). These events are passed through a full detector
simulation and subjected to the same
reconstruction and event selection procedure applied to the data. The distribution
of the selected events in the Dalitz plot is then used to
determine the total reconstruction and selection efficiency. 
The MC sample $e^+ e^- \to D_s^{*+} X$, where $D_s^{*+} \to \gamma \Ds$, used to 
compute this efficiency consists of 27.4 $\times 10^6$ generated events for 
$\Ds \to \pip \pim \pip$ 
and 4.2 $\times 10^6$ for $\Ds \to \Kp \Km \pip$.
The Dalitz plot is divided into small cells and the efficiency 
distribution is fitted with
a second-order polynomial in two dimensions. The efficiency is found to be almost uniform
as a function of the $\pip\pim$ invariant mass with an average value of $\approx$ 1.6 \%.
This low efficiency is mainly due to the likelihood ratio selection: it is 18.0\% without this cut.

The experimental resolution as a function of the $\pip \pim$ mass has been computed as the 
difference between MC generated and reconstructed mass. It increases from 1.0 to 2.5 \mevcc from the
$\pip \pim$ threshold to 1.0 \gevcc.

\section{Dalitz Plot Analysis}
An unbinned maximum likelihood fit is performed
on the distribution of events 
in the Dalitz plot to determine the relative amplitudes and phases 
of intermediate resonant and nonresonant states.
The likelihood function is:

\begin{equation}
\begin{split}
\mathcal{L} = \prod_{\rm events} [x(m) \cdot \eta(m_1^2,m_2^2){ \frac{\sum_{i,j} c_i c_j^* A_i A_j^*}
{\sum_{i,j} c_i c_j^* I_{A_i A_j^*}}} \\
 +  (1-x(m))\frac{\sum_{i} |k_i|^2 B^2_i}
{\sum_{i} |k_i|^2 I_{B^2_i}}]
\end{split}
\end{equation}

\noindent where:
\begin{itemize}
\item $m_1^2$ and $m_2^2$ are the squared $\pip \pim$ effective masses;
\item $x(m)$ is the mass-dependent fraction of signal, defined as $x(m) = \frac{G(m)}{G(m)+P(m)}$.
Here $G(m)$ and $P(m)$ represent the
Gaussian and the linear function used to fit the $\pip \pim \pip$ mass spectrum.
\item $\eta(m^2_1,m^2_2)$ is the efficiency, parametrized with a two-dimensional
second order polynomial.
\item $A_i, B_i$ describe signal and background amplitude contributions respectively.
\item $k_i$ are real factors describing the structure of background. They are computed by fitting the sideband regions.
\item $I_{A_i A_j^*}=\int A_i A_j^* \eta(m^2_1,m^2_2) dm_1^2 dm_2^2$ and $I_{B^2_i}$ are the normalization
integrals for signal and background respectively. The products of efficiency and amplitudes are normalized using a numerical 
integration over the Dalitz plot.
\item $c_i$ are complex coefficients allowed to vary during the fit procedure.
\end{itemize}

The efficiency-corrected fraction due to the resonant or nonresonant contribution $i$
is defined as follows:
\begin{eqnarray}
f_i = \frac {|c_i|^2 \int |A_i|^2 dm_1^2 dm_2^2}
{\sum_{j,k} c_j c_k^* \int A_j A_k^* dm_1^2 dm_2^2}.
\end{eqnarray}
The $f_i$ values do not necessarily add to 1 because of interference effects.
The uncertainty on each $f_i$ is evaluated by propagating the 
full covariance matrix obtained
from the fit.

The phase of each amplitude ({\it i.e.} the phase of the corresponding $c_i$) is measured with respect to 
the $f_2(1270) \pip$ amplitude. 
Each $\mathcal{P}$-wave and $\mathcal{D}$-wave amplitude $A_i$ is represented by the product of a complex 
Breit-Wigner ($BW(m)$) and a real angular term:
\begin{eqnarray}
A = BW(m) \times T (\Omega).
\end{eqnarray}
\noindent where $m$ is the $\pip \pim$ mass. 
The Breit-Wigner function includes the Blatt-Weisskopf form 
factors~\cite{dump}. The angular terms $T (\Omega)$ are described in
Ref.~\cite{pdg}. 

For the $\pip \pim$ $\mathcal{S}$-wave amplitude, we use a different approach because:
\begin{itemize}
\item Scalar resonances have large uncertainties. In addition, the existence of some 
states needs confirmation.
\item Modelling the $\mathcal{S}$-wave as a superposition of Breit-Wigners 
is unphysical since it leads to a violation of unitarity 
when broad resonances overlap. 
\end{itemize}
To overcome these problems, we use a Model-Independent Partial Wave Analysis  
introduced by the Fermilab E791 Collaboration~\cite{Aitala:2005yh}: instead of 
including the $\mathcal{S}$-wave amplitude as a superposition of relativistic Breit-Wigner functions, we divide 
the $\pip\pim$ mass spectrum into 29 slices and we parametrize the $\mathcal{S}$-wave by an 
interpolation between the 30 endpoints in the complex plane: 
\begin{eqnarray}
A_{\mathcal{S}-{\rm wave}}(m_{\pi\pi}) = {\rm Interp}(c_k(m_{\pi\pi})e^{i\phi_k(m_{\pi\pi})})_{k=1,..,30}.
\end{eqnarray}
The amplitude and phase of each endpoint are free parameters. The width of each slice is tuned to get 
approximately the same number of $\pip \pim$ combinations ($\simeq 13179 \times 2/29$).
Interpolation is implemented by a Relaxed Cubic Spline~\cite{cern1}. The phase is not constrained 
in a specific range in order to allow the spline to be a continuous function. 

The background shape is 
obtained by fitting the \Ds sidebands.
In this fit, resonances are assumed to be incoherent, {\it i.e.} are represented
by Breit-Wigner intensity terms only. A good representation of the background includes
contributions from $K^0_S$, $\rho^0(770)$ and three ad-hoc scalar resonances
with free parameters.
  
Resonances are included in sequence, keeping only those having a fractional significance
greater than two standard deviations. 

\begin{figure*}[t]
\begin{center}
\includegraphics[width=12cm]{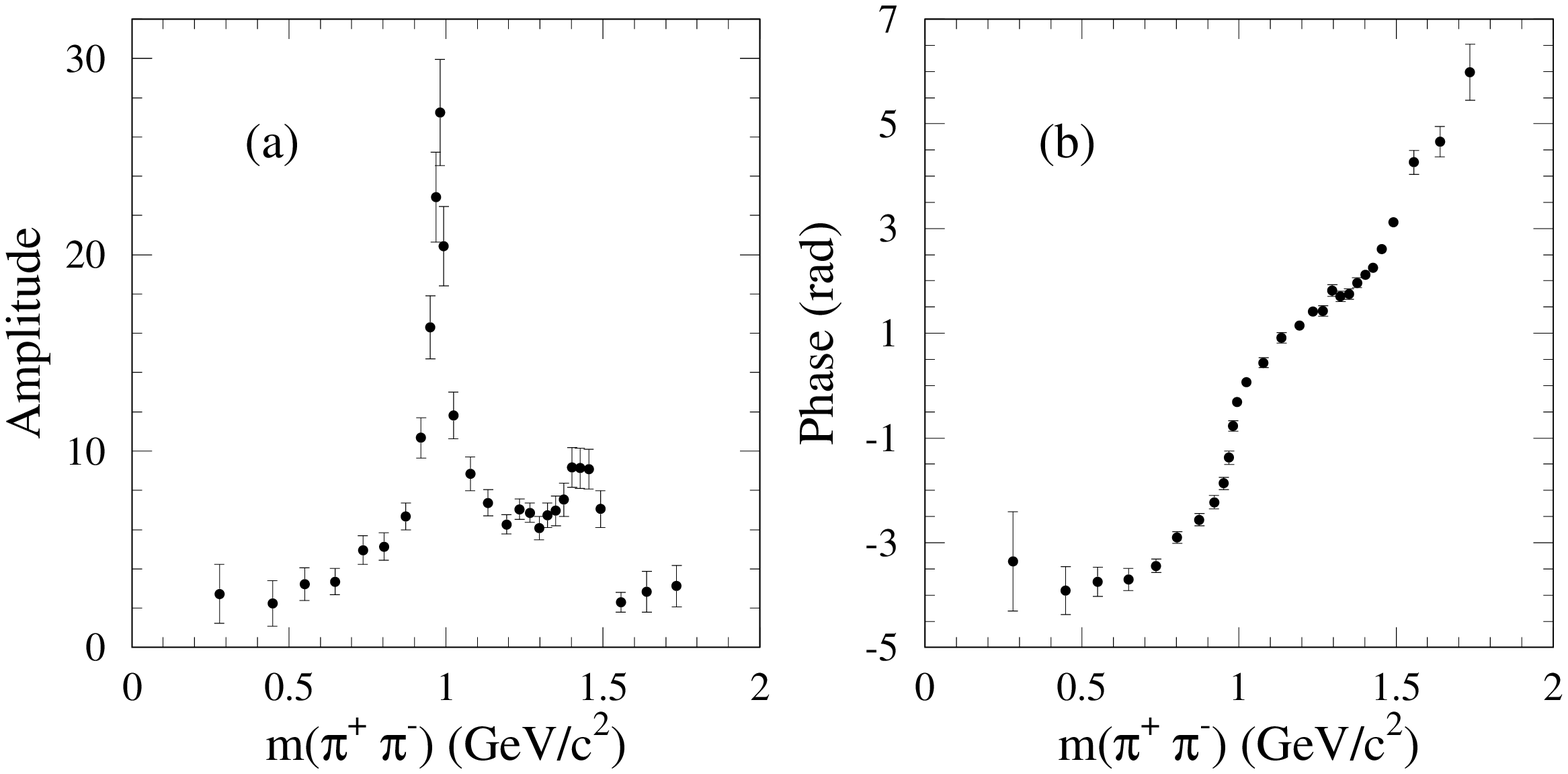}
\includegraphics[width=12cm]{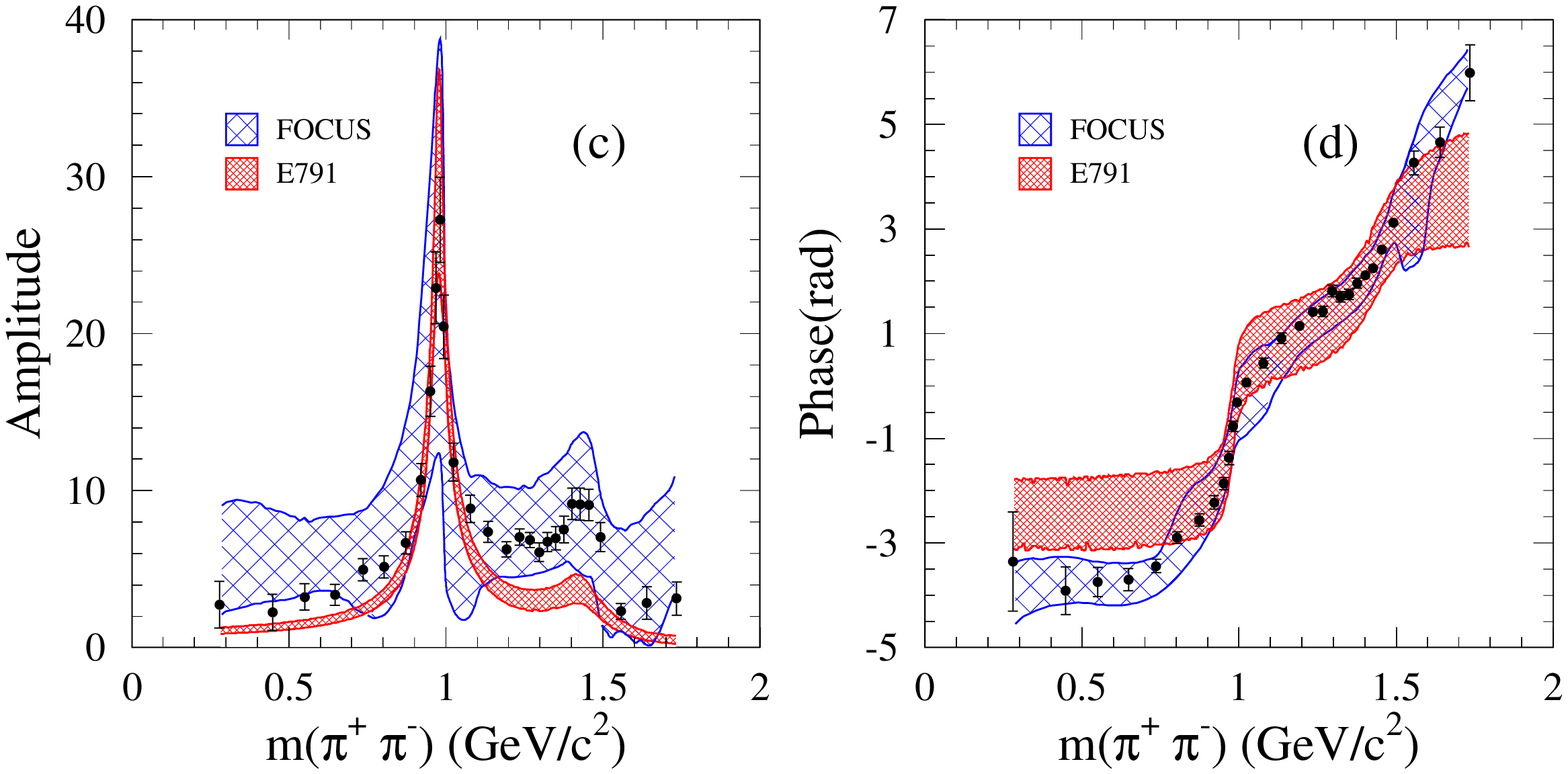}
\caption{(a) $\mathcal{S}$-wave amplitude extracted from the best fit, (b) corresponding $\mathcal{S}$-wave phase, (c) $\mathcal{S}$-wave amplitude compared to  the FOCUS and E791 amplitudes, (d) $\mathcal{S}$-wave phase compared to the FOCUS and E791 phases. Errors are statistical only.}
\label{fig:fig_2}
\end{center}
\end{figure*}

\section{Results}
The fit results (fractions and phases)
are summarized in Table~\ref{tab:fit_pipipi}. The resulting $\mathcal{S}$-wave $\pip \pim$ amplitude and phase is shown in 
Fig.~\ref{fig:fig_2}(a),(b) and is given numerically in Table~\ref{tab:bin_pipipi}. Fig.~\ref{fig:fig_2}(c),(d) show
a comparison with the resulting $\mathcal{S}$-wave from the E791 experiment, which performed a Dalitz 
plot analysis using an isobar model~\cite{e791_ds}, and the FOCUS experiment, which made use of the K-matrix formalism~\cite{focus}.
In the two figures, the two bands have been obtained by propagating the measurement errors and assuming no correlations.
This assumption may influence the calculation of the uncertainties on the phases and amplitudes which are different
in the two experiments.

\begin{table}[!htb]
\centering
\caption{Results from the $D_s^+ \to \pip \pim \pip$ Dalitz plot analysis. The table reports the fit fractions, amplitudes and
phases. Errors are statistical and systematic respectively.}
\label{tab:fit_pipipi}
\begin{tabular}{@{}cc@{}cc@{}}
\hline
\hline
Decay Mode\rule{0pt}{9pt}      & Decay fraction(\%)&Amplitude&Phase(rad)\\
\hline
$f_2(1270)\pip$\rule{0pt}{9pt} &$10.1\pmpm1.5\pmpm1.1$& 1.(Fixed)            &   0.(Fixed)          \\
$\rho(770)\pip$                &$1.8\pmpm0.5\pmpm1.0$&$0.19\pmpm0.02\pmpm0.12$&$1.1\pmpm0.1\pmpm0.2$\\
$\rho(1450)\pip$               &$2.3\pmpm0.8\pmpm1.7$&$1.2\pmpm0.3\pmpm1.0$&$4.1\pmpm0.2\pmpm0.5$\\
$\mathcal{S}$-wave              &$83.0\pmpm0.9\pmpm1.9$&{\bf Table~\ref{tab:bin_pipipi}}&{\bf Table~\ref{tab:bin_pipipi}}\\
\hline
Total                           &$97.2\pmpm3.7\pmpm3.8$&&\\
$\chi^2/NDF$\rule{0pt}{9pt}     &$\frac{437}{422-64}$ = 1.2&                                 &\\
\hline
\hline
\end{tabular}
\end{table}

\begin{figure*}[!htb]
\begin{center}
\includegraphics[width=12cm]{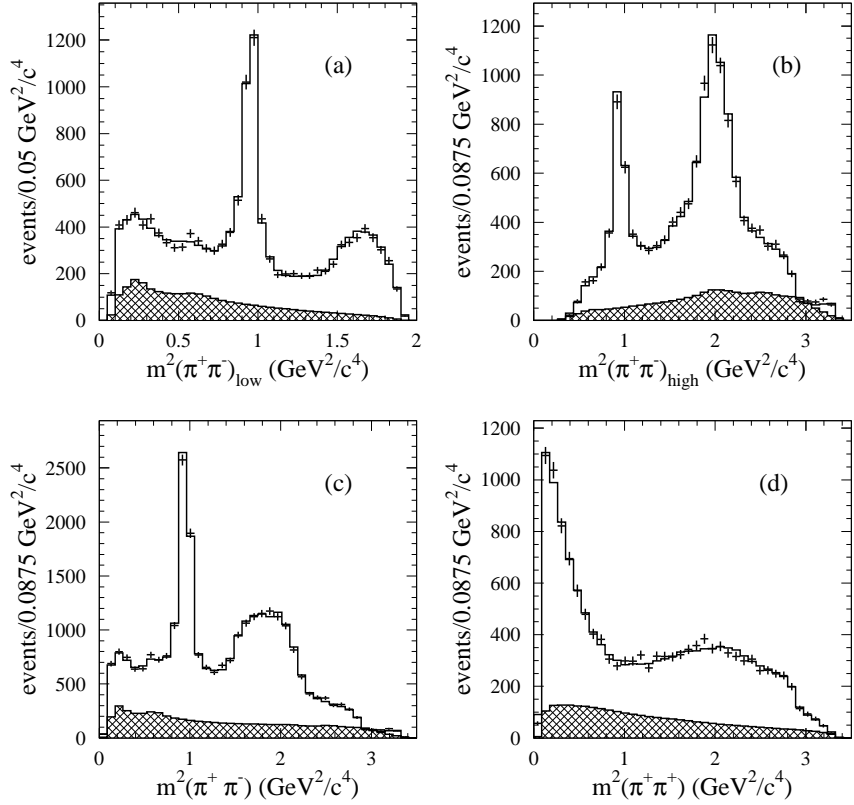}
\caption{Dalitz plot projections (points with error bars) and fit results (solid histogram). 
(a) $m^2(\pip \pim)_{{\rm low}}$, (b) $m^2(\pip \pim)_{{\rm high}}$, (c) total $m^2(\pip \pim)$, (d) $m^2(\pip \pip)$. 
The hatched histograms show the background distribution.}
\label{fig:fig_3}
\end{center}
\end{figure*}

The Dalitz plot projections 
together with the fit
results are shown in Fig.~\ref{fig:fig_3}.
Here we label with $m^2(\pip \pim)_{{\rm low}}$ and 
$m^2(\pip \pim)_{{\rm high}}$ the lower and higher values of the two $\pip \pim$ mass combinations.

\begin{figure*}[!htb]
\begin{center}
\includegraphics[width=15.6cm]{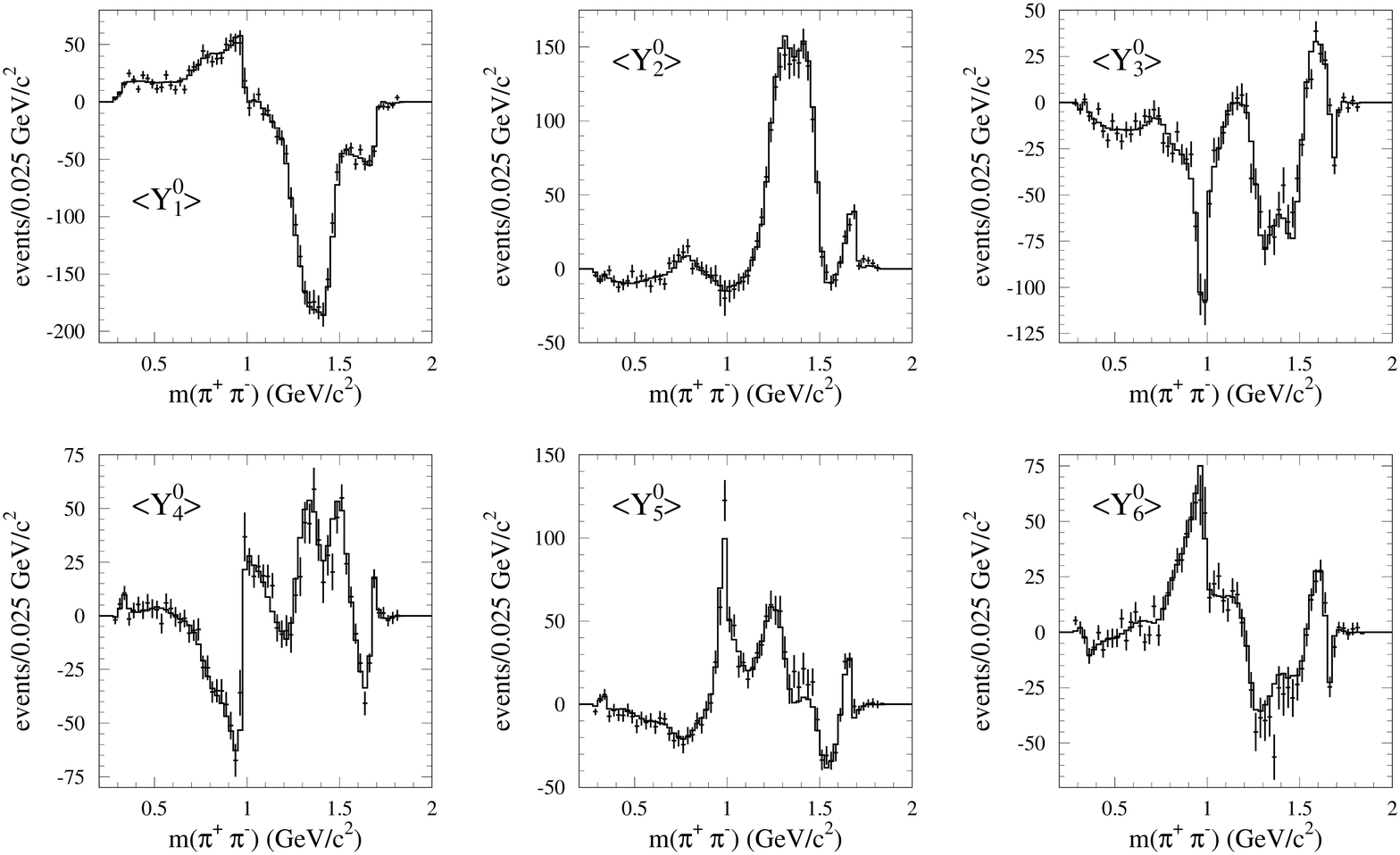}
\caption{Unnormalized spherical harmonic moments $\left <Y^0_L \right >$ 
as a function of $\pip \pim$ effective mass. The data are presented with error bars, the
histograms represent the fit projections.}
\label{fig:fig_4}
\end{center}
\end{figure*} 
The fit projections are obtained by generating a large number of phase space MC
events~\cite{cern2}, weighting by the fit likelihood function, and normalizing the weighted sum to 
the observed number of events.
There is good agreement 
between data and fit projections. 
Further tests of the fit quality are performed using unnormalized $Y^0_L$ moment projections
onto the $\pip \pim$ axis as functions of the helicity angle $\theta$, which 
is defined as the 
angle between the $\pim$ and the $D^+_s$
in the $\pip \pim$ rest frame (or $\pip$ for $D^-_s$) (two combinations per event).
The $\pip \pim$ mass distribution is then weighted by
the spherical harmonic $Y_L^0(\cos \theta)$ ($L=1-6$).
The resulting distributions of the $\left<Y^0_L \right>$ are shown in 
Fig.~\ref{fig:fig_4}. 
A straightforward interpretation of these distributions is difficult, due to
reflections originating from the symmetrization. However, the squares of the spin amplitudes appear in even moments, while interference terms appear in odd moments.
 
The fit produces a good representation of the data for all 
projections.
The fit $\chi^2$ is computed by dividing the Dalitz plot into 30$\times$30 cells with 422 cells
having entries. We obtain $\chi^2/NDF= 437/(422-64)=1.2$. 
The $\chi^2$ is also calculated using an adaptive binning  with an average number of events per cell $\simeq$35 
(\mbox{$\chi^2/NDF = 365/(391-64)=1.1$}), obtaining a $\chi^2$ probability of 7.2\%.

Attempts to include
other resonant contributions, such as $\omega(782)$ or $f_2'(1525)$, do not improve the fit quality.
MC simulations have been performed in order to validate the method and test
for possible multiple solutions. 

The results from the Dalitz plot analysis can be summarized as follows:
\begin{itemize}
\item{} The decay is dominated by the $\Ds \to (\pip \pim)_{\mathcal{S}-{\rm wave}} \pip$ contribution.
\item{} The $\mathcal{S}$-wave shows, in both amplitude and phase, the expected behavior for the $f_0(980)$ resonance.
\item{} The $\mathcal{S}$-wave shows further activity, in both amplitude and phase, in the regions of the $f_0(1370)$ and $f_0(1500)$ 
resonances.
\item{} The $\mathcal{S}$-wave is small in the $f_0(600)$ region, indicating that this resonance has a small coupling to $s \bar s$.
\item{} There is an important contribution from $\Ds \to f_2(1270) \pip$ whose size is in 
agreement with that reported by FOCUS, but a factor two smaller than that reported by E791. 
This is the largest contribution 
in charm decays from a spin-2 resonance.
\item{} We observe a similar trend for the $\mathcal{S}$-wave amplitude and phase between the three experiments. 
Our results agree better 
(within uncertainties) with the results from FOCUS than those from E791.
\end{itemize} 
Our results may be compared with different measurements of the $\pi \pi$ amplitude and phase from many other sources. For a recent 
review, see~\cite{klempt}.

Systematic uncertainties on the fitted fractions are evaluated in different ways:
\begin{itemize}
\item{} The background parametrization is performed using the information from the lower/higher sideband only or both
sidebands.
\item{} The Blatt-Weisskopf barrier factors have a single parameter $r$ which we take to be 1.5 $({\rm GeV}/c)^{-1}$
and which has been varied between 0 and 3 $({\rm GeV}/c)^{-1}$.
\item{} Results from fits which give equivalent Dalitz plot descriptions and similar sums of fractions
(but worse likelihood) are considered.
\item{} The likelihood cut is relaxed but the mass cut on the $\pip \pim \pip$ is narrowed in order to obtain a similar purity.
\item{} The purity of the signal, the resonance parameters and the efficiency coefficients are
varied within their statistical errors.
\item{} The $\rho(770)$ and $\rho(1450)$ parametrization is modified according to the Gounaris-Sakurai model~\cite{gusak}.
\item{} The number of steps used to describe the  $\mathcal{S}$-wave has been varied by $\pm 2$. 
\end{itemize}

\section{Branching Fraction}

Since the two \Ds decay channels (1) and (3) have similar topologies,
the ratio of branching fractions is expected to have a reduced
systematic uncertainty. We therefore select events from the two \Ds decay modes using similar selection criteria
for the $D^{*+}_s$ selection and for the likelihood test. For this measurement, a looser likelihood cut is used. 

The ratio of branching fractions is evaluated as:
\begin{eqnarray}
BR = \frac{\sum_{x,y} \frac{N_1(x,y)}{\epsilon_1(x,y)}}{\sum_{x,y}\frac{N_0(x,y)}{\epsilon_0(x,y)}},
\end{eqnarray}
where $N_i(x,y)$ represents the number of events measured for channel $i$,
and $\epsilon_i(x,y)$ is the corresponding efficiency in a given Dalitz 
plot cell $(x,y)$. For this calculation each Dalitz plot was divided into 50$\times$50 cells.

To obtain the yields and measure the relative branching fractions,
the $\pip\pim\pip$ and $\Kp \Km \pip$ mass distributions
are fit assuming a double 
Gaussian signal and linear background where all the parameters are floated. 
Systematic uncertainties, summarized in Table~\ref{syst}, take into account uncertainties from MC statistics and from the selection criteria used. 

The resulting ratio is:
\begin{eqnarray}
\frac{ \BR(D_s^+ \to \pip \pim \pip)}{ \BR(D_s^+ \to \Kp \Km \pip)} = 0.199 \pm 0.004 \pm 0.009
\end{eqnarray}
\noindent consistent, within one standard deviation, with the PDG~\cite{pdg} value: $0.265 \pm 0.041 \pm 0.031$. 
It is also consistent with a recent measurement from CLEO~\cite{cleo_br}: $0.202 \pm 0.011 \pm 0.009$.

The study of the $D_s^+ \to \Kp \Km \pip$ decay can give new information on the $K \bar K$ $\mathcal{S}$-wave.
This information together with the results reported in this analysis will enable new measurements of
the $f_0(980)$ couplings to $\pi\pi/K \bar K$.  

\section{Conclusions}
A Dalitz plot analysis of approximately 13,000 $\Ds \to \pip \pim \pip$ has been performed. 
The fit measures  fractions and phases for quasi-two-body decay modes. The amplitude and phase of the 
$\pip \pim$ $\mathcal{S}$-wave is extracted in a model-independent way for the first time.
We also measure with high precision the $\BR(D_s^+ \to \pip \pim \pip)/ \BR(D_s^+ \to \Kp \Km \pip)$
ratio.

\begin{table}[!htb]
\begin{center}
\caption{Amplitude and phase of the $\pim\pip$ $\mathcal{S}$-wave amplitude determined with the MIPWA fit described in the text. The first error is statistical while the second is systematic.}
\label{tab:bin_pipipi}
\begin{tabular}{c c r@{}c@{}l r@{}c@{}l}
\hline
\hline
Interval\rule{0pt}{11pt} & Mass (\gevcc)&\multicolumn{3}{c}{Amplitude}& \multicolumn{3}{c}{Phase(radians)}\\
\hline
1\rule{0pt}{9pt}   & 0.28 & 2.7 $\pm$& \, 1.5 \, & $\pm$ 2.4 & -3.4 $\pm$& \, 1.0 \,& $\pm$ 1.3\\
2   & 0.448 & 2.2 $\pm$& \, 1.2 \, & $\pm$ 1.3 & -3.9 $\pm$& \, 0.5 \,& $\pm$ 0.4\\
3   & 0.55 & 3.2 $\pm$& \, 0.8 \, & $\pm$ 1.1 & -3.7 $\pm$& \, 0.3 \,& $\pm$ 0.3\\
4   & 0.647 & 3.3 $\pm$& \, 0.7 \, & $\pm$ 0.9 & -3.7 $\pm$& \, 0.2 \,& $\pm$ 0.3\\
5   & 0.736 & 5.0 $\pm$& \, 0.7 \, & $\pm$ 1.1 & -3.4 $\pm$& \, 0.1 \,& $\pm$ 0.2\\
6   & 0.803 & 5.1 $\pm$& \, 0.7 \, & $\pm$ 0.8 & -2.9 $\pm$& \, 0.1 \,& $\pm$ 0.2\\
7   & 0.873 & 6.7 $\pm$& \, 0.7 \, & $\pm$ 0.7 & -2.6 $\pm$& \, 0.1 \,& $\pm$ 0.3\\
8   & 0.921 & 10.7 $\pm$& \, 1.0 \, & $\pm$ 0.9 & -2.2 $\pm$& \, 0.1 \,& $\pm$ 0.2\\
9   & 0.951 & 16.3 $\pm$& \, 1.6 \, & $\pm$ 1.2 & -1.9 $\pm$& \, 0.1 \,& $\pm$ 0.2\\
10   & 0.968 & 22.9 $\pm$& \, 2.3 \, & $\pm$ 1.5 & -1.4 $\pm$& \, 0.1 \,& $\pm$ 0.1\\
11   & 0.981 & 27.2 $\pm$& \, 2.7 \, & $\pm$ 1.6 & -0.8 $\pm$& \, 0.1 \,& $\pm$ 0.2\\
12   & 0.993 & 20.4 $\pm$& \, 2.0 \, & $\pm$ 0.9 & -0.3 $\pm$& \, 0.1 \,& $\pm$ 0.2\\
13   & 1.024 & 11.8 $\pm$& \, 1.2 \, & $\pm$ 0.5 & 0.1 $\pm$& \, 0.1 \,& $\pm$ 0.2\\
14   & 1.078 & 8.8 $\pm$& \, 0.9 \, & $\pm$ 0.3 & 0.4 $\pm$& \, 0.1 \,& $\pm$ 0.1\\
15   & 1.135 & 7.4 $\pm$& \, 0.7 \, & $\pm$ 0.3 & 0.9 $\pm$& \, 0.1 \,& $\pm$ 0.1\\
16   & 1.193 & 6.3 $\pm$& \, 0.5 \, & $\pm$ 0.2 & 1.1 $\pm$& \, 0.1 \,& $\pm$ 0.1\\
17   & 1.235 & 7.0 $\pm$& \, 0.5 \, & $\pm$ 0.3 & 1.4 $\pm$& \, 0.1 \,& $\pm$ 0.1\\
18   & 1.267 & 6.9 $\pm$& \, 0.5 \, & $\pm$ 0.3 & 1.4 $\pm$& \, 0.1 \,& $\pm$ 0.1\\
19   & 1.297 & 6.1 $\pm$& \, 0.6 \, & $\pm$ 0.6 & 1.8 $\pm$& \, 0.1 \,& $\pm$ 0.1\\
20   & 1.323 & 6.7 $\pm$& \, 0.6 \, & $\pm$ 0.5 & 1.7 $\pm$& \, 0.1 \,& $\pm$ 0.1\\
21   & 1.35 & 7.0 $\pm$& \, 0.8 \, & $\pm$ 0.6 & 1.8 $\pm$& \, 0.1 \,& $\pm$ 0.2\\
22   & 1.376 & 7.5 $\pm$& \, 0.8 \, & $\pm$ 0.7 & 2.0 $\pm$& \, 0.1 \,& $\pm$ 0.1\\
23   & 1.402 & 9.2 $\pm$& \, 1.0 \, & $\pm$ 0.9 & 2.1 $\pm$& \, 0.1 \,& $\pm$ 0.1\\
24   & 1.427 & 9.1 $\pm$& \, 1.0 \, & $\pm$ 0.9 & 2.3 $\pm$& \, 0.1 \,& $\pm$ 0.2\\
25   & 1.455 & 9.1 $\pm$& \, 1.0 \, & $\pm$ 1.6 & 2.6 $\pm$& \, 0.1 \,& $\pm$ 0.1\\
26   & 1.492 & 7.0 $\pm$& \, 0.9 \, & $\pm$ 1.1 & 3.1 $\pm$& \, 0.1 \,& $\pm$ 0.2\\
27   & 1.557 & 2.3 $\pm$& \, 0.5 \, & $\pm$ 0.7 & 4.3 $\pm$& \, 0.2 \,& $\pm$ 0.4\\
28   & 1.64 & 2.8 $\pm$& \, 1.1 \, & $\pm$ 1.3 & 4.7 $\pm$& \, 0.3 \,& $\pm$ 0.7\\
29   & 1.735 & 3.1 $\pm$& \, 1.1 \, & $\pm$ 2.3 & 6.0 $\pm$& \, 0.5 \,& $\pm$ 1.4\\
\hline
\hline
\end{tabular}
\end{center}
\end{table}

\begin{table}[!htbp]
\centering
\caption{Summary of systematic uncertainties on the $\BR(D_s^+ \to \pip \pim \pip)/ \BR(D_s^+ \to \Kp \Km \pip)$ ratio.\\}
\begin{tabular}{lc} 
\hline
\hline
Source  & Systematic uncertainties (\%) \\ \hline
 MC statistics                   &  0.9  \\ 
 $\Delta m$ cut                   &  1.5  \\ 
 Likelihood cut                  &  2.6  \\ 
 Particle identification      &  3.0  \\ 
\hline
 Total                           &  4.3  \\ 
\hline
\hline
\end{tabular}
\label{syst}
\end{table}

\section{Acknowledgements}

We are grateful for the 
extraordinary contributions of our \pep2\ colleagues in
achieving the excellent luminosity and machine conditions
that have made this work possible.
The success of this project also relies critically on the 
expertise and dedication of the computing organizations that 
support \babar.
The collaborating institutions wish to thank 
SLAC for its support and the kind hospitality extended to them. 
This work is supported by the
US Department of Energy
and National Science Foundation, the
Natural Sciences and Engineering Research Council (Canada),
the Commissariat \`a l'Energie Atomique and
Institut National de Physique Nucl\'eaire et de Physique des Particules
(France), the
Bundesministerium f\"ur Bildung und Forschung and
Deutsche Forschungsgemeinschaft
(Germany), the
Istituto Nazionale di Fisica Nucleare (Italy),
the Foundation for Fundamental Research on Matter (The Netherlands),
the Research Council of Norway, the
Ministry of Education and Science of the Russian Federation, 
Ministerio de Educaci\'on y Ciencia (Spain), and the
Science and Technology Facilities Council (United Kingdom).
Individuals have received support from 
the Marie-Curie IEF program (European Union) and
the A. P. Sloan Foundation.

\vfill

\renewcommand{\baselinestretch}{1}

\end{document}